\documentclass[acmsmall,screen,nonacm]{acmart}

\usepackage{amsfonts} 
\usepackage{amsmath}  
\usepackage{proof} 
\usepackage{hyperref} 

\usepackage{tikz}
\usetikzlibrary{positioning}

\title{Bridging Syntax and Semantics of Lean Expressions in E-Graphs}

\author{Marcus Rossel}
\email{marcus.rossel@tu-dresden.de}
\affiliation{
	\institution{Chair for Compiler Construction, TU Dresden}
	\country{Germany}
}

\author{Andr\'es Goens}
\email{a.goens@uva.nl}
\affiliation{
	\institution{University of Amsterdam}
	\country{Netherlands}
}

\begin{document}

\maketitle

Interactive theorem provers, like \emph{Isabelle/HOL}, \emph{Coq} and \emph{Lean}, have expressive languages that allow the formalization of general mathematical objects and proofs.
In this context, an important goal is to reduce the time and effort needed to prove theorems.
A significant means of achieving this is by improving proof automation.
We have implemented an early prototype of proof automation for equational reasoning in Lean by using equality saturation.\footnote{\url{https://github.com/marcusrossel/lean-egg}}
To achieve this, we need to bridge the gap between Lean's expression semantics and the syntactically driven e-graphs in equality saturation.
This involves handling bound variables, implicit typing, as well as Lean's \emph{definitional equality}, which is more general than syntactic equality and involves notions like $\alpha$-equivalence, $\beta$-reduction, and $\eta$-reduction.
In this extended abstract, we highlight how we attempt to bridge this gap, and which challenges remain to be solved.
Notably, while our techniques are partially unsound, the resulting proof automation remains sound by virtue of Lean's proof checking.

\section{Overview of the Proof Tactic}

\begin{figure}[H]
	\begin{center}
		\scalebox{0.94}{
			\begin{tikzpicture}[every node/.style={inner sep = 1mm, minimum height = 7mm}]

				\node (tactic) [draw, rounded corners] {\texttt{egg} Tactic};
				\coordinate[left = 5mm of tactic] (c0);
				\coordinate[right = 5mm of tactic] (c1);
				\coordinate[above = 5mm of c1] (c1a);
				\coordinate[below = 5mm of c1] (c1b);
				\coordinate[right = 4cm of c1] (c2);
				\coordinate[above = 5mm of c2] (c2a);
				\coordinate[below = 5mm of c2] (c2b);
				\node[right = 0 of c1a] (goal) [draw, rounded corners] {Goal Equation};
				\node[right = 0 of c1b] (theorems) [draw, rounded corners] {Equational Theorems};
				\node[right = 0 of c2a] (goal-exprs) [draw, rounded corners] {Goal Expressions};
				\node[right = 0 of c2b] (rewrites) [draw, rounded corners] {Rewrites};
				\node[right = 35mm of c2] (egg) [draw, rounded corners] {egg};
				\node[below = 5mm of egg] (explanation) [draw, rounded corners] {Explanation};
				\node[below = 5mm of tactic] (proof) [draw, rounded corners] {Proof};
				\draw [->] (c0) to (tactic);
				\draw [->] (tactic) to (goal);
				\draw [->] (tactic) to (theorems);
				\draw [->] (goal) to (goal-exprs);
				\draw [->] (theorems) to (rewrites);
				\draw [->] (goal-exprs) to (egg);
				\draw [->] (rewrites) to (egg);
				\draw [->] (egg) to (explanation);
				\draw [->] (explanation) to (proof);
				\draw [->] (proof) to (tactic);
			\end{tikzpicture}
		}
	\end{center}
	\setlength{\abovecaptionskip}{1.5mm}
	\caption{Overview of the \texttt{egg} proof tactic.}
	\label{fig:proof-tactic-overview}
\end{figure}
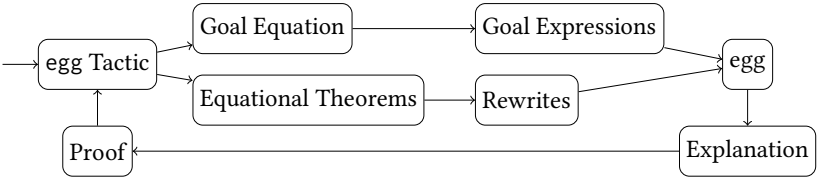

We start with a brief overview of how we use equality saturation~\cite{tate2009-equality-saturation} with Lean~\cite{demoura2021lean} by connecting it to egg~\cite{willsey2021-egg}.
The setup, sketched in Figure~\ref{fig:proof-tactic-overview}, is both based on and extends the tactic described in~\cite{DBLP:journals/pacmpl/KoehlerGBGTS24}.

\paragraph*{Expressions}

Lean is based~\cite{carneiro2024lean4lean} on the calculus of constructions~\cite{coquand1988calculus}, which is an expressive, dependently-typed $\lambda$ calculus.
In this calculus, data types and their values, as well as propositions and proofs are all terms of the same expression language:

\vspace{-2.5mm}

\[
	\mathit{e} \; ::= \; \hat{n} \; | \; e\,e \; | \; \lambda e, e \; | \; \forall e, e \quad \quad \text{where} \; n \in \mathbb{N}
\]

\vspace{0.5mm}

It contains abstraction at the term and type level ($\lambda$ and $\forall$), based on de Bruijn indices ($\hat{n}$), and function application.\footnote{This is a simplification of Lean's expression language, to simplify the presentation of key aspects in this extended abstract.}
The expression appearing directly after a binder provides the type of the bound variable $\hat{0}$.
All terms in Lean are lowered to expressions of this language at \emph{elaboration} (compile) time.
Thus, to use Lean terms in egg, we directly encode this expression language in egg.
When the specific representation of expressions is not important, we often write $\lambda$-expressions with named variables and without explicit binder types in subsequent sections.
Also, we often use the terms ``term'' and ``expression'' interchangeably.

\paragraph*{Proof Tactics}

Proofs in Lean are seldom written as explicit expressions of the given $\lambda$ calculus.
Instead, they are built by calling a sequence of \emph{proof tactics}, which are small \mbox{(meta-)programs} that construct the relevant proof terms.
For example, the $\mathsf{rfl}$ tactic solves proof goals of the form $x = y$, if $x$ and $y$ are (definitionally) equal.\footnote{We explain the notion of \emph{definitional equality} in Section~\ref{sec:semantics-of-eq}.}
The tactic $\mathsf{rewrite\,[eqn_1, eqn_2, ...]}$ rewrites the proof goal by applying each given equational theorem once.
We implement \emph{our} proof automation as a tactic $\mathsf{egg\,[eqn_1, eqn_2, ...]}$, which solves proof goals of the form $x = y$ by having egg find a sequence of given rewrites which transform $x$ to $y$.

\paragraph*{Proof Generation}

One of the key properties of theorem provers like Lean is the LCF architecture~\cite{DBLP:conf/praapr/Milner72}, where proofs are checked for correctness by a small \emph{trusted kernel} of code assumed to be correct.
Thus, proof automation tools can play fast and loose without compromising soundness of the system, as incorrect proofs will not be accepted by the kernel.
This comes at the cost of automated proof tools, like egg, having to generate proof witnesses for their results.
Additionally, these witnesses need to be converted into a format which Lean understands.
Luckily, egg supports generating \emph{explanations} which can be expressed as linear sequences of rewrites.
These rewrites are then (more or less) easily converted into valid proofs in Lean.

\section{The (Lack of) Problems with Binders}

The choice of directly using Lean's expression language in egg may raise immediate concerns to some readers.
Scalably representing $\lambda$ calculus in e-graphs is notoriously difficult due to the presence of binders.
Binders make it necessary to consider expressions \emph{in context} instead of purely syntactically.
For example, when using de Bruijn indices, the expression $\hat{1}$ refers to a free variable in $\lambda \hat{1}$, but to a bound variable in $\lambda \lambda \hat{1}$.
The problem persists when using named variables where $x$ refers to a free variable in $\lambda y, x$ --- but to a bound variable in $\lambda x, x$.
Thus, directly representing variables in e-graphs causes three classes of problems:

\subsection{Free Variable Rebinding}

The first class of problems is well-known in $\lambda$ calculus.
When a term is rewritten, the resulting substitutions may cause free variables to turn into bound variables -- that is, they may be \emph{invalidly captured} by a binder.
For example, if we apply the rewrite $?x \mapsto (\lambda\, ?x)\,0$ to the expression $\lambda\,\hat{0}$, we get the non-equivalent $\lambda ((\lambda\, \hat{0})\, 1)$.\footnote{In this example, even though $\hat{0}$ appears under a binder, it is a free variable with respect to the left-hand side of the rewrite.}
Notably, free variables can also become invalidly ``un-captured''.
For example, if we apply the rewrite $\lambda (\lambda\, ?x)\, 0 \mapsto \lambda\, ?x$ to the open term $\lambda (\lambda\, \hat{4})\, 0$, we get the non-equivalent $\lambda\, \hat{4}$.
Here, $\hat{4}$ would need to be shifted to $\hat{3}$ to retain its semantics.
Both of these problems can be avoided by having every rewrite be a dynamic rewrite which checks whether a given matched pattern variable refers to free variables, and if so, shifts them as needed.
Checking for free variables can be done efficiently by using an e-class analysis.
Shifting variables, on the other hand, is rather involved as it needs to be performed on the results of e-matching which are e-classes instead of expressions.
A procedure for this is discussed in Section~\ref{sec:subst}.

\subsection{Invalid Matching}

The second class of problems is \emph{invalid matching} of pattern variables against (1) syntactically equal but semantically different variables and (2) ``locally bound'' variables.
An example of (1) occurs when the pattern term $\lambda (\lambda\, ?x)\, ?x$ is matched against $\lambda (\lambda\, \hat{0})\, \hat{0}$.
Here, the matched variables $\hat{0}$ refer to different variables, even though they are syntactically equal.
An example of (2) occurs when the rewrite $(\lambda\, ?x)\, 1 \mapsto\, ?x$ is applied to $(\lambda\, \hat{0})\, 1$, which produces the non-equivalent term $\hat{0}$.
The reason this fails is that the given rewrite is valid only when $?x$ is not bound.\footnote{All rewrites where the trigger pattern contains binders are valid only if pattern variables are not matched against bound variables of those binders.}
As $?x$ appears under one binder in the trigger pattern, any matched variable with index $< 1$ is bound in the context of the trigger pattern (we call this \emph{locally bound}) and therefore invalid.
Invalid matching is easily avoided with dynamic rewrites which check for occurrences of (1) and (2) by a traversal of the trigger pattern.
If any pattern variable maps to a variable and appears under multiple different binders, we have case (1).
If a pattern variable at binder depth $d$ maps to a variable whose index is smaller than $d$, we have case (2).
In either case, the rewrite is aborted.

\subsection{Bound Variable Aliasing}

A more fundamental problem with direct representation of variables is \emph{invalid aliasing}.
This occurs when different expressions use the same bound variable name, which is necessarily the case for de Bruijn indices.
For example, in an e-graph with expressions $\lambda\,\hat{0} + 1$ and $\lambda\,\neg \hat{0}$, the same e-node for $\hat{0}$ is used by both expressions.
Thus, any term shown equivalent to $\hat{0}$ in one expression automatically becomes equivalent in both.
This can easily become a problem in implicitly-typed languages like Lean's expression language.
For example, if we rewrite $\hat{0}$ to $1 \cdot \hat{0}$ in the first expression, then as a result of aliasing the second one represents the term $\lambda \, \neg(1 \cdot \hat{0})$, which is ill-typed.

We believe the problem of bound variable aliasing to be partially solvable by adding explicit type annotations to all terms.
This directly solves aliasing of variables with different types, as they become syntactically distinct.
It, thus, suffices to show that aliasing is a non-issue for variables of the same type.
While we do not have proof of this, we have an argument which relies on the fact that expressions which are equivalent to variables must be equivalent in all contexts.
Thus, one only needs to consider what happens to the variables of terms which become equivalent to another variable, as they can be invalidly un-/captured.
We think that these scenarios are all benign, though.
A major caveat to this approach comes from the fact that adding explicit type annotation might be non-trivial.
As types in Lean can be parameterized, type annotations themselves may need to contain variables.
Thus, our argument may only apply to settings where all type are unparameterized.
We have not found this to be an issue in practice, in any of the examples we have seen so far.

\section{Semantics of Equality}
\label{sec:semantics-of-eq}

Even if we ignore the problems of binders, Lean's expression language does not naturally collaborate with e-graphs' syntax-driven approach to rewrites.
In an e-graph, two expressions are equal a priori only if they are syntactically equal.
In Lean, expressions are equal a priori if they satisfy a certain \emph{definitional equality} relation ($\equiv$).
For example, $(\lambda \, x, x + 0) \, 1 = 1$ holds by reflexivity in Lean, but is not provable without rewrites in an e-graph.
This difference requires consideration as it means that we might not identify a goal as being proven during equality saturation, simply because we could only reach a term which was definitionally but not syntactically equal to the goal.
It also has effects on how rewrites can be applied in Lean as compared to an e-graph.
If we have a rewrite like $(\lambda \, l, l + 1) \, l_1 = (\lambda \, l, l + 1) \, l_2$, we cannot use it to prove $l_1 + 1 = l_2 + 1$ by equality saturation, as the rewrite does not match any goal term.
To overcome these differences, we add rewrites which allow conversion between different definitionally equal representations of the same expression.
Some rules of definitional equality are already intrinsic to e-graphs and do not require rewrites.
Namely, those which express that definitional equality is an equivalence relation which is a congruence with respect to application, and abstraction.
All other forms of definitional equality require various techniques outlined below.

\paragraph{Normalization}

Some definitional equalities can be achieved by eliminating certain syntactic constructs from all expressions -- that is, by partially normalizing expressions.
For example, Lean has a notion of $\mathsf{let}$-expressions in its language with a corresponding definitional equality rule.
We avoid this rule by completely reducing $\mathsf{let}$-expressions (this is called $\zeta$-reduction) before equality saturation.
Thus, the expression language used in the e-graph has no concept of $\mathsf{let}$-expressions, making their definitional equality rules irrelevant.

\paragraph{Erasure}

Lean has a rule known as \emph{proof irrelevance} which states that all proofs of the same proposition are definitionally equal.
We overapproximate this rule and make \emph{all} proofs equal in an e-graph by erasing proof terms completely.
That is, if we encounter a proof term while encoding a Lean expression to be used in an e-graph, we replace it by some constant symbol $\epsilon$.
Reconstructing these erased terms then requires additional work during proof generation in Lean.
We are so far unsure if there are cases where erased proof terms cannot be completely reconstructed.

\paragraph{Dynamic Rewrites}

The definitional equality rules $\beta$ and $\eta$ establish relationships between application, abstraction and substitution.

\vspace{-3.5mm}

\[
	\infer[(\beta)]{(\lambda \alpha, e)\,e' \equiv e[\hat{0} \mapsto e']}{} \quad \quad
	\infer[(\eta)]{(\lambda \alpha, e \, \hat{0}) \equiv e}{\quad\hat{0} \not\in \operatorname{fvars}(e)\;}
\]

\vspace{0.5mm}

When working with de Bruijn indices, both $\beta$- and $\eta$-reduction necessitate shifting indices of free variables.
Namely, in the case of $\eta$, the expression $\lambda\, e\, \hat{0}$, is only equivalent to the $\eta$-reduced $e$ if all free variables in $e$ are shifted down by $1$.
This is necessary as we remove one binder above $e$.
For $\beta$-reduction, the reduced expression $f[\hat{0} \mapsto e]$ also requires shifting all free variables in $f$ down by $1$ as we again remove one binder above $f$.
Additionally, all free variables in $e$ need to be shifted up by the number of binders appearing above the corresponding occurrence of $\hat{0}$ in $f$.
This is necessary to compensate for the fact that we are (potentially) adding binders above $e$.
Thus, to encode $\beta$- and $\eta$-reduction we need dynamic rewrites.
We start by implementing shifting of variables which can be generalized to substitution of variables with constant expressions.\footnote{We mean \emph{constant} in the sense that the substituted expression will be inserted as is.}
Notably, we cannot simply implement this function over expressions, but must implement it on e-classes, as e-matching does not yield expressions but only e-classes.
The required function $\mathsf{subst}$ can thus be described as follows.
Let $\sigma : \mathbb{N} \times \mathbb{N} \to \mathit{expr}$ be a substitution mapping from a variable index and binder depth to an expression.
Let $g$ be an e-graph containing an e-class $c$.
Then $\mathsf{subst}(g, c, \sigma)$ extends $g$ with an e-class $s$ such that for every expression $e$ represented by $c$, $s$ represents $\sigma'(e)$.
We use $\sigma'$ to denote the lifting of $\sigma$ to apply to expressions with the binder depth being derived starting at $c$.
Using $\mathsf{subst}$ we can define $\eta$- and $\beta$-reduction entirely based on the choice of $\sigma$.\footnote{The last case in $\sigma_\beta$ returns an e-class instead of just a single expression. This does not match our definition of a substitution, but $\mathsf{subst}$ can easily be adapted to handle this, too.}

\vspace{-3.5mm}

\begin{align*}
	\mathsf{eta}(g, c)                                       & := \mathsf{subst}(g, c, \sigma_{\pm}(-1))
	\\
	\mathsf{beta}(g, c, \mathit{arg})                        & := \mathsf{subst}(g, c, \sigma_\beta(\mathit{arg}))
	\\
	\sigma_{\pm}(\mathit{off})(\mathit{idx}, \mathit{depth}) & :=
	\begin{cases}
		\widehat{\mathit{idx} + \mathit{off}} & \text{if} \; \mathit{idx} > \mathit{depth} \\
		\widehat{\mathit{idx}}                & \text{otherwise}
	\end{cases}
	\\
	\sigma_\beta(\mathit{arg})(\mathit{idx}, \mathit{depth}) & :=
	\begin{cases}
		\widehat{\mathit{idx} - 1}                                    & \text{if} \; \mathit{idx} > \mathit{depth} \\
		\widehat{\mathit{idx}}                                        & \text{if} \; \mathit{idx} < \mathit{depth} \\
		\mathsf{subst}(g, \mathit{arg}, \sigma_{\pm}(\mathit{depth})) & \text{if} \; \mathit{idx} = \mathit{depth} \\
	\end{cases}
	\\
\end{align*}

We can then construct dynamic rewrites for $\beta$- and $\eta$-reduction by e-matching on their respective pattern expressions, performing $\mathsf{beta}$ or $\mathsf{eta}$ on the e-class of the matched variable, and finally forming the union of the resulting e-class with the pattern expression's e-class.
The implementation of $\mathsf{subst}$ is discussed in Section~\ref{sec:subst}.

\section{Substitution on E-Classes}
\label{sec:subst}

Substitution on e-classes is another notoriously difficult problem with different solutions.

\begin{enumerate}
	\item Make substitution an explicit constructor of the expression language and introduce rewrites which encode its semantics (with respect to other constructors), as described in~\cite{willsey2021-egg}.
	\item Perform substitution by extracting a single expression $e$ from the e-class and performing substitution on $e$ before adding it back into the e-graph, as described in ~\cite{DBLP:phd/ethos/Koehler22}.
	\item Traverse the subgraph of the target e-class and construct an equivalent subgraph where all relevant nodes are substituted.
\end{enumerate}

As (1) introduces subtleties with respect to correctly handling rewrite rules and makes scaling harder by increasing the e-graph~\cite{DBLP:phd/ethos/Koehler22}, and (2) lacks completeness by potentially never extracting all represented expressions, we choose (3) for our implementation of the $\mathsf{subst}$ function.
Recall that our goal is for $\mathsf{subst}(g, c, \sigma)$ to extend the e-graph $g$ with an e-class $s$ such that for every expression $e$ represented by e-class $c$, $s$ represents $\sigma'(e)$ where $\sigma : \mathbb{N} \times \mathbb{N} \to \mathit{expr}$ is a substitution mapping from a variable index and binder depth to an expression.
Thus, our approach is to traverse the subgraph of $c$ and in the process construct a new e-class $s$ where variable nodes are substituted according to $\sigma$.
Traversal of an e-graph rooted at a given e-class really means two kinds of traversal.
First, we must traverse all e-nodes in the e-class and, second, we must traverse all children of each e-node.
For this, we use a basic recursive depth-first traversal which additionally iterates over each e-node of a given e-class on visit.
As e-graphs can contain cycles, we make sure to avoid infinite recursion by remembering which e-classes have already been visited.
Problems then only arise once we try to replace variable e-nodes with new e-nodes.

\paragraph*{E-Class Creation}

The fundamental problem with replacing e-nodes is that it can only be done bottom-up.
For example, let $c_1$ be an e-class containing an e-node $n_1$ with child e-class $c_2$ containing an e-node $n_2$, etc, until finally $n_k$ is a terminal e-node.
Then consider how to construct the substituted e-class $s_1$ which represents the same terms as $c_1$, but with $n_k$ replaced by some other term $m_k$.
Intuitively, we might first construct a new empty e-class for $s_1$, then (recursively) construct the substituted e-class $s_2$ and finally add e-node $n_1$ with new child $s_2$ into $s_1$.
Unfortunately, we cannot construct empty e-classes and instead can only create new e-classes starting at leaf nodes.
Thus, we instead need to first traverse all the way down to $n_k$, create the new e-class for $m_k$, and then start constructing the new e-classes $s_i$ bottom-up.
While this is arguably just as simple as the ``intuitive'' approach, it breaks when the e-graph has a cycle.
Namely, when we reach an e-class $c$ which has been visited before, $c$ may not yet have an associated substituted e-class, as potentially none of its child e-classes have completed substitution yet.
Thus, we run into a situation akin to a deadlock, where in order to construct the substituted version $s$ of $c$, we need to construct a substituted e-node which itself depends on $s$.
Deciding how to continue the traversal in that case is the key aspect of our substitution algorithm.

\paragraph*{Cycle Breaking}

To break cyclic dependencies encountered during e-class creation, we rely on an important cycle breaking property of e-graphs.
Every cycle in an e-graph contains at least one e-node $n$ where each child e-class of $n$ is either (1) the root of an acyclic subgraph or (2) the root of a subgraph satisfying the cycle breaking property.
Intuitively, this corresponds to the fact that an e-graph can never contain an e-class which does not represent any (finite) term.
This property holds trivially for an empty e-graph and is preserved under adding expressions.
Thus, in a proof, it suffices to show that it is also preserved under e-class merging.
Based on the cycle breaking property, our substitution algorithm resolves cyclic dependencies using an approach similar to data driven scheduling in process networks \cite{parks1995bounded}.

Let $n$ be an e-node for which we are trying to construct a substitute by recursively calling $\mathsf{subst}$ on its child e-classes.
If a child $c_i$ has already been visited but does not yet have a substitute e-class (that is, if we have encountered a cycle in the e-graph), then $\mathsf{subst}$ returns a failure value for $c_i$.
We record the failed child e-classes in a set $\mathcal{C}$.
If $\mathcal{C} = \emptyset$, then we can immediately construct a substitute for $n$ based on the substituted children.
If $\mathcal{C} \ne \emptyset$, we record the fact that $n$ has not yet been substituted and is waiting for the e-classes in $C$.
The waiting e-nodes are then reconsidered whenever a new substitute e-class is created.
Specifically, we check whether any waiting e-node's dependencies have all been substituted, and, if so, construct the e-node's substitute.
Thus, by the cycle breaking property, all waiting e-nodes will at some point be substituted.

A major caveat to our algorithm is that we do not yet know how to properly handle justifications (used for explanations).
Thus, while the algorithms is perfectly suitable for applications which trust egg, it sometimes falls short for our proof automation purposes by losing crucial rewrite steps needed for proof reconstruction.

\section{Conclusion}

The semantics of Lean expressions rely on context-dependent constructs like binders and definitional equality, while rewriting in equality saturation is chiefly syntactic.
This difference can lead to rewrites applying when they should not (by invalid matching and bound variable aliasing), as well as rewrites not applying when they should (by definitional but not syntactic equality).
Various approaches can be used to bridge this gap.
Normalization of expressions, term erasure and dynamic rewrites resolve different definitional equalities.
Dynamic rewriting based on free variable analysis inhibits invalid matching and variable capture.
Additionally, explicit type annotations could restrict invalid bound variable aliasing.
A common denominator between many dynamic rewrites is their need for substitution, for which we have shown one approach by means of e-graph traversal.
Most of the given approaches are implemented as part of a proof tactic in Lean which uses egg as an automated procedure for equational reasoning.
Preliminary tests show that these techniques are promising for this use case.

\bibliography{main}

\end{document}